\documentclass{iau}
\usepackage{graphicx}
\usepackage{natbib}
\usepackage{aas_macros}

\title[Sunquakes and starquakes] 
{Sunquakes and starquakes}

\author[Alexander G. Kosovichev]   
{Alexander G. Kosovichev$^{1,2,3}$}

\affiliation{$^1$Hansen Experimental Physics Laboratory, Stanford University, Stanford, CA 94305, USA \\
email: {\tt AKosovichev@solar.stanford.edu} \\[\affilskip]
$^2$Big Bear Solar Observatory/NJIT, Big Bear City, CA 92314, USA \\[\affilskip]
$^3$Crimean Astrophysical Observatory, Nauchny, Crimea 98409, Ukraine}

\pubyear{2014}
\volume{301}  
\pagerange{1--4}
\jname{Precision Asteroseismology}
\editors{J.A. Guzik, W.J. Chaplin, G. Handler \& A. Pigulski, eds.}
\begin{document}

\maketitle

\begin{abstract}
In addition to well-known mechanisms of excitation of solar and
stellar oscillations by turbulent convection and instabilities, the
oscillations can be excited by an impulsive localized force caused by
the energy release in solar and stellar flares. Such oscillations
have been observed on the Sun (`sunquakes'), and created a lot
of interesting discussions about physical mechanisms of the
impulsive excitation and their relationship to the flare physics.
The observation and theory have shown that most of a sunquake's
energy is released in high-degree, high-frequency p modes. In
addition, there have been reports on helioseismic observations of
low-degree modes excited by strong solar flares. Much more powerful
flares observed on other stars can cause `starquakes' of
substantially higher amplitude. Observations of such oscillations
can provide new asteroseismic information and also constraints
on mechanisms of stellar flares. I discuss the basic properties
of sunquakes, and initial attempts to detect
flare-excited oscillations in \textit{Kepler} short-cadence data.
\keywords{Sun: flares, oscillations; stars: flares, oscillations}
\end{abstract}

\firstsection 
\section{Introduction}

\textit{Kepler} observations have led to a discovery that stellar
flares representing impulsive powerful energy releases occur not only
in M-type dwarfs (UV Ceti-type variables) but also in a wide range of
A-F type stars
\citep{Balona2012,Maehara2012,Walkowicz2011}. Previously, it was
believed the F- and A-type stars do not have flaring activity. The
discovery of super-flares on solar-type stars raised questions about
the possibility of such flares on the Sun. These results triggered new
debates about the physical mechanism of stellar flares and the
relationship to the dynamo mechanism. Recent observations of solar
flares from the Solar Dynamics Observatory (SDO) have found that the
white-light flares are often accompanied by excitation of acoustic
oscillations (p modes) in the solar interior, so-called `sunquakes'.
However, not all flares reveal the sunquakes. The relationship of the
interior and atmospheric response to the energy release and optical
emission is not understood, and is currently a subject of detailed
studies in heliophysics. The flare-excited oscillations are mostly
observed as local seismic waves, and while the theory predicts that
the global acoustic waves are also excited, their amplitude is
significantly lower than the amplitude of stochastically excited
oscillations, therefore, the detection reports have been
controversial. However, in the case of significantly more powerful
stellar flares the impact on the star's surface is much greater. This
can lead to excitation of the global low-degree oscillations to
significantly higher amplitudes than on the Sun. Our preliminary study
of the available short-cadence (SC) data provides indications of such
`starquakes'. However, a statistical study for a large sample of stars
and longer observing intervals are needed.

\begin{figure}[t]
\begin{center}
\includegraphics[width=0.7\textwidth]{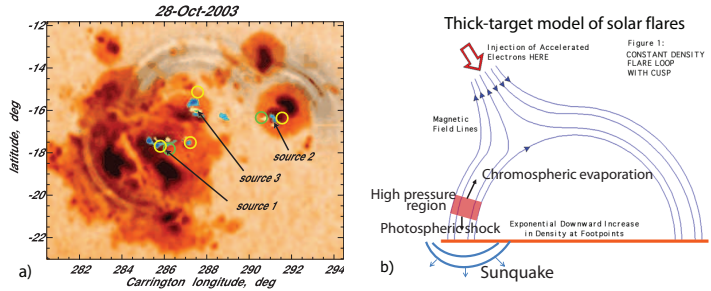}
\caption{a) Observations of
acoustic wavefronts (projected onto a sunspot image) during a large
solar flare. Yellow and green circles show places of the hard X-ray
and $\gamma$-ray emissions, and blue patches are Doppler-shift signals
$>1$ km\,s$^{-1}$ from the localized flare impacts \citep{Kosovichev2006};
b) illustration of excitation of
`sunquakes' in the `thick-target' model of solar flares.} 
\label{fig1}
\end{center}
\end{figure}

\section{Solar flares and sunquakes}

\begin{figure}[t]
\begin{center}
\includegraphics[width=\textwidth]{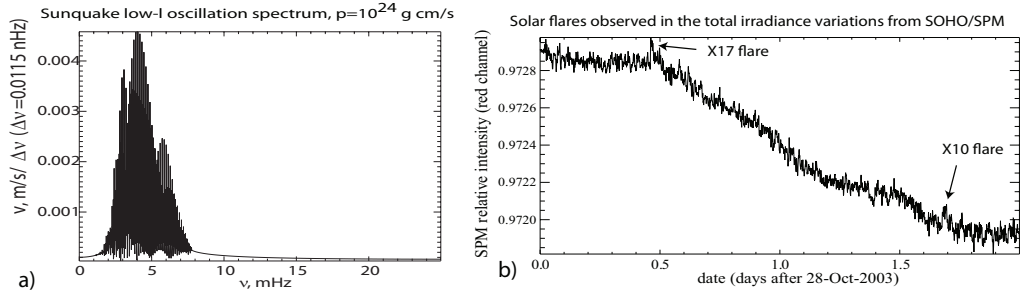}
\caption{a) The theoretical spectrum
of the flare excited oscillations has a maximum around the acoustic
cut-off frequency $\sim $5~mHz \citep{Kosovichev2009}; (b) The signal of two large solar flares observed in
the total irradiance measurements (red channel) with the SPM
instrument on the  \textit{SOHO} spacecraft.}
\label{fig2}
\end{center}
\end{figure}

Recent observations of solar flares on the NASA space missions
\textit{SOHO} and \textit{SDO} revealed that some flares result in
excitation of acoustic oscillations in the solar interior, dubbed
`sunquakes'. The oscillations are observed in the Doppler shift and
intensity variations as expanding circular waves
(Fig.~\ref{fig1}a). They are excited due to momentum and energy
impulse in the solar photosphere, which is also observed in solar
flares, but the mechanism of this impact is unknown. The expanding
waves represent high-degree p modes; however, the stellar oscillation
theory predicts that the global low-degree modes are also
excited. Their detection has been reported from a statistical analysis
of the correlation between the flare soft X-ray signals (observed by
the \textit{GOES} satellites) and the total solar irradiance
measurements from the space observatory \textit{SOHO}
\citep{Karoff2008}. However, there was no unambiguous detection of the
whole-Sun oscillations caused by individual flare events. Indeed, the
oscillation theory predicts that the amplitudes of the low-degree
modes of sunquakes are significantly lower than the amplitudes of
stochastically excited oscillations (Fig.~\ref{fig2}a)
\citep{Kosovichev2009}.

The solar observations show that the sunquake events are mostly
associated with compact or confined flares that do not produce coronal
mass ejections and are characterized by the energy release in compact
magnetic configurations in the lower atmosphere. Such flares usually
produce white-light or continuum optical emission, and they do not
necessarily have strong X-ray emission. The origin of such flare
energy release is mysterious, and the subject of intensive
investigation. These flares and sunquakes challenge the standard
`thick-target' flare model, which assumes that most of the flare
energy is released in the form of high-energy particles due to magnetic
reconnection processes in the coronal plasma. The particles travel along
the magnetic field lines and heat the upper chromosphere to high
temperature, creating a high-pressure region, which expands, producing
a plasma eruption (`chromospheric evaporation') and a
downward-traveling radiative shock (Fig.~\ref{fig1}b). This shock can
reach the photosphere and excite acoustic oscillations. However, the
numerical simulations of the flare hydrodynamics show that the energy
of the shock may be not sufficient to explain the observed
oscillations, and also the photospheric impact is often observed at
the beginning of the flare impulsive phase or even during the
pre-heating phase, well before the maximum of the hard and soft X-ray
emissions that are supposedly produced by the particle interaction
with the higher chromosphere. Nevertheless, if a similar energy
release mechanism works on other stars then the amplitudes of the
flare-excited oscillations can be expected to be several orders of
magnitude stronger than on the Sun.

\section{Initial analysis of starquakes}

Observations of stellar flares from \textit{Kepler} provide a unique
opportunity to investigate the relationship between the flare energy
release and oscillations. The analysis of \textit{Kepler} data by
\citet{Balona2012}, \citet{Maehara2012} and \citet{Walkowicz2011} has
led to the discovery of a large number of flaring stars of various
spectral classes, from M- to A-class. While large flares on M-K stars
were known from ground-based observations, the discovery of similar
flares on F-A stars is surprising. However, most of the flare data
were obtained with long cadence, and do not provide sufficient
resolution to estimate the flare amplitude and duration of the various
phases of the energy release, or to investigate the seismic
response of the stars. For understanding the mechanisms of flares and
the impact of these flares, it is important to obtain more
short-cadence data for flaring stars.

The previous \textit{Kepler} observations allowed us to select stars
that showed multiple flares and with good quality oscillation data
(when short-cadence runs are available). We visually inspected the
light curves of the previously detected flaring stars, as well as the
database of the ground observations.

\begin{figure}[t]
\begin{center}
\includegraphics[width=\textwidth]{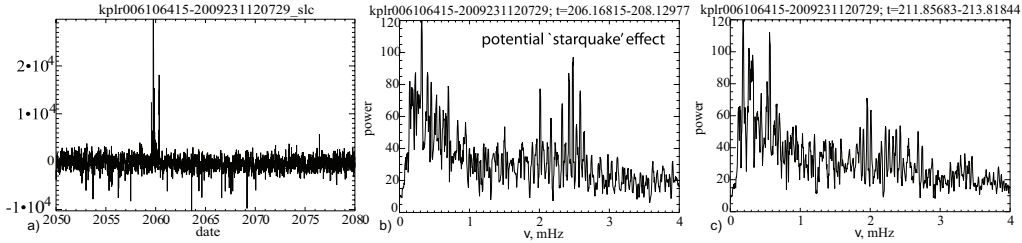}
\caption{Illustration of a `starquake' candidate obtained from the SC
\textit{Kepler} observations of KIC 6106415: (a) the \textit{Kepler} light curve of a
stellar flare. The oscillation power spectra: (b) calculated for a
2-day period immediately after the flare, and (c) taken during the star quiet
period, more than 5 days after the flare.} 
\label{fig3}
\end{center}
\end{figure}

The flare brightness signals observed by \textit{Kepler} (as
illustrated in Fig.~\ref{fig3}a) are much stronger than those on the
Sun (Fig.~\ref{fig2}a).  Thus, the photospheric impacts can be much
stronger, and the flare oscillations can be observed in the
\textit{Kepler} data.  In Figure~\ref{fig3} we illustrate the
potential detection of a `starquake' event obtained from the SC
\textit{Kepler} observations of KIC 6106415.  The \textit{Kepler}
light curve of this stellar flare is shown in panel (a). The
oscillation power spectrum calculated for a 2-day period immediately
after the flare (Fig.~\ref{fig3}b) shows an enhancement of the
p-mode amplitudes at $\sim 2.6$~mHz, closer to the acoustic cut-off
frequency for this star ($\sim 3.5$~mHz), compared to the
oscillation spectrum taken during the star's quiet period
(Fig.~\ref{fig3}c), more than 5 days after the flare (the so-called
"numax" parameter at a frequency of $\sim 2.26$~mHz, e.g., see
Chaplin et al., 2013). Of course, more statistical studies are
needed to confirm this result.

Oscillations associated with stellar flares have previously been
observed in optical and X-ray emissions with  periods ranging
from a few seconds to tens of minutes, and have usually been interpreted as
transient oscillations of flaring loops in stellar atmospheres
\citep{Contadakis2012,Jakimiec2012,Koljonen2011,Contadakis2012a,Qian2012}. However,
the atmospheric transient oscillations quickly decay, within 1-2 hours
after a flare impulse \citep{Bryson2005}, while the stellar
oscillation modes will live for much longer, depending on the damping
mechanism in stellar envelopes. Also, the maximum amplitude of the
oscillations modes excited by flares is expected to be in the range of
3-20 min, close to the cut-off period of acoustic oscillations. We
cannot rule out that oscillations with longer periods, corresponding
to g modes and mixed modes, are also excited. Nevertheless, we plan
the initial search for the flare seismic responses in the frequency
range around the acoustic cut-off frequency.

\section{Discussion}

The initial results from the \textit{Kepler} mission have demonstrated
its tremendous capability for studying stellar activity and
oscillations. In particular, the previous observing programs have
discovered that flares, during which stellar brightness dramatically
increases for short periods of time, are common in both cool and hot
stars.  Previously, stellar flares were associated with active M-type
stars, the UV Ceti variables, thought to be similar to solar flares,
which represent a sudden release of magnetic energy accumulated in
the coronal part of sunspot regions in the form of high-energy
particles, which heat the lower atmosphere. However, the white-light
emission in solar flares is rare, and there are only few cases when it
has been unambiguously observed in the intergrated Sun-as-a-star
observations. The stellar flares can be four orders-of-magnitude more
powerful. This might be due to bigger sunspot regions generated by a more
efficient dynamo process, because many of the flaring stars rotate
faster than the Sun. However, there is an alternative point of view
that the flares may be due to  interactions with close companions, `hot
Jupiters'. The discovery of similar flares on hot A-type stars with a
very shallow outer convection zone and without strong magnetic fields
\citep{Balona2012} raises additional problems with the dynamo origin
of the flare energy.



\end{document}